# Privacy-Preserving Quantum Two-Party Geometric Intersection

**Wen-Jie Liu[1, 2, *], Yong Xu[2], James C. N. Yang[3], Wen-Bin Yu[1, 2] and Lian-Hua Chi[4]**

**Abstract：** Privacy-preserving computational geometry is the research area on the intersection of the domains of secure multi-party computation (SMC) and computational geometry. As an important field, the privacy-preserving geometric intersection (PGI) problem is when each of the multiple parties has a private geometric graph and seeks to determine whether their graphs intersect or not without revealing their private information. In this study, through representing Alice's (Bob's) private geometric graph $G_A$ ( $G_B$ ) as the set of numbered grids $S_A$ ( $S_B$ ), an efficient privacy-preserving quantum two-party geometric intersection (PQGI) protocol is proposed. In the protocol, the oracle operation $O_A$ ( $O_B$ ) is firstly utilized to encode the private elements of $S_A = (a_0, a_1, \cdots, a_{M-1})$ ( $S_B = (b_0, b_1, \cdots, b_{N-1})$ ) into the quantum states, and then the oracle operation $O_f$ is applied to obtain a new quantum state which includes the XOR results between each element of $S_A$ and $S_B$. Finally, the quantum counting is introduced to get the amount ( $t$ ) of the states $\left| a_i \oplus b_j \right\rangle$ equaling to $\left| 0 \right\rangle$, and the intersection result can be obtained by judging $t > 0$ or not. Compared with classical PGI protocols, our proposed protocol not only has higher security, but also holds lower communication complexity.

## 1. Introduction

The problem of privacy-preserving computational geometry is an important research area on the intersection of the domains of secure multi-party computation (SMC) [Oleshchuk and Zadorozhny (2007)] and computational geometry [Preparata and Shamos (2012)]. It focuses on how cooperative users can use their own private geometric information as inputs in collaborative computing in the distributed systems, and they can obtain the correct results while ensuring their privacy. Since the privacy-preserving computational geometry is firstly proposed by Atallah et al. [Atallah and Du (2001)], the other researchers have drawn extensive attention on some related problems, such as point inclusion [Troncoso-Pastoriza, Katzenbeisser, Celik et al. (2007); Luo, Huang and Zhong (2007)], geometric intersection [Erlebach, Jansen and Seidel (2005); Pawlik, Kozik, Krawczyk et al. (2013)], nearest points or closest pair [Li and Ni (2002); Tao, Yi, Sheng et al. (2010)], and convex hull [Huang, Luo and Wang (2008); Löffler and van Kreveld (2010); Assarf, Gawrilow, Herr et al. (2017)], which have been applied to many important military and commercial fields.

---

[1] Jiangsu Engineering Center of Network Monitoring, Nanjing University of Information Science & Technology, Nanjing 210044, China.

[2] School of Computer and Software, Nanjing University of Information Science and Technology, Nanjing 210044, China.

[3] Department of Computer Science and Information Engineering, National Dong Hwa University, Hualien 974, Taiwan.

[4] Department of Computer Science and Information Technology, La Trobe University, VIC 3086, Australia.

[*] Corresponding Author: Wen-Jie Liu. Email: wenjiel@163.com.

Consider the following scenario, two countries A and B intend to build a railway in an offshore area. Before the completion of the railway, the construction route is confidential. In order to prevent future collisions of trains, countries A and B hope to determine if there are any two disjoint routes without revealing their own routes, and to negotiate with the location of the intersection. The above problem is a typical application of privacy-preserving geometric intersection (PGI). Different from the protocols based on circuit evaluation schemes, recently Qin et al. [Qin (2014)] proposed the Lagrange multiplier method to solve the intersection of the two private curves, and this method is suitable for solving general geometry intersection problems. On the other way, some researchers tried to study the geometric problems in three dimensional space [Li, Wu, Wang et al. (2014)]. However, most of these classical solutions are based on computational complexity assumptions, and they cannot ensure the participants' privacy under the attack of quantum computation.

Fortunately, quantum cryptography can provide the unconditional security, which is guaranteed by some physical principles of quantum mechanics, to resist against such impact. In additional, quantum parallelism makes it possible to greatly speed up solving some specific computational tasks, such as large-integer factorization [Shor (1994)] and database search [Grover (1996)]. With quantum mechanics utilized in the information processing, many important research findings are presented in recent decades, such as quantum key distribution (QKD) [Bennett and Brassard (1984)], quantum key agreement (QKA) [Liu, Chen, Ji et al. (2017); Liu, Xu, Yang et al. (2018)], quantum secure direct communication [Liu, Chen, Ma et al. (2009); Liu, Chen, Liu (2016), Liu and Chen (2016)], quantum private comparison [Liu, Liu, Liu et al. (2014); Liu, Liu, Chen et al. (2014); Liu, Liu, Wang et al. (2014)], and quantum sealed-bid auction (QSBA) [Naseri (2009); Liu, Wang, Ji et al. (2014); Liu, Wang, Yuan et al. (2016)], and deterministic remote state preparation [Liu, Chen, Liu et al. (2015); Qu, Wu, Wang et al. (2017)]. These findings have shown the potential power in either the efficiency improvements or the security enhancements.

In this study, we pay attention to the PGI problem: Alice owns a private geometric graph $G_A$, Bob has the other geometric graph $G_B$, and they want to determine whether these two graphs intersect without revealing any private information to each other. By utilizing some specific oracle operations and quantum counting algorithm, we propose an efficient privacy-preserving quantum two-party geometric intersection (PQGI) protocol. The rest of this paper is organized as follows, the PQGI protocol is proposed in Sect. 2, and the correctness, security and efficiency analysis of PQGI protocol are discussed in Sect. 3, while the conclusion is drawn in the last section.

## 2. Preliminaries

Before introducing the procedures of PQGI protocol, we firstly make some definitions of PGI problem and PQGI protocol. Without loss of generality, we suppose there are two parties, i.e., Alice and Bob, and the formal definitions are given as below.

### 2.1. The problems

**Problem 1 (Privacy-preserving point inclusion):** There are two parties, Alice has a point $p_A$, and Bob has a geometric graph $G_B$. They want to decide whether $p_A \in G_B$ without revealing to each other anything more than what can be inferred from that answer.

**Problem 2 (Privacy-preserving two-party geometric intersection):** Two parties Alice, Bob own the private geometric graphs $G_A$, $G_B$, respectively, and decide whether $G_A \cap G_B \neq \varnothing$ without disclosing their respective private information.

As a point can be viewed as a special geometric graph whose area is small enough to be one dot, Problem 1 is a typical case of Problem 2. In the study, we only consider the geometric intersection of problem 2.

## 2.2. the definition of PQGI

In order to solve Problem 2, the private geometric graph can be represented as the set of grids in the area of the graph (suppose these girds are divided sufficiently), then the intersection of two geometric graphs is transformed into the intersection of two sets. Without loss of generality, we suppose Alice and Bob have a private geometric graph $G_A$ and $G_B$ on the plane, and they divide and number the whole plane into $R$ grids (Here $R$ is a large enough integer), then Alice's and Bob's graphs can be denoted as $S_A = (a_0, a_1, \cdots, a_{M-1})$, $S_B = (b_0, b_1, \cdots, b_{N-1})$, respectively (shown in Fig. 1).

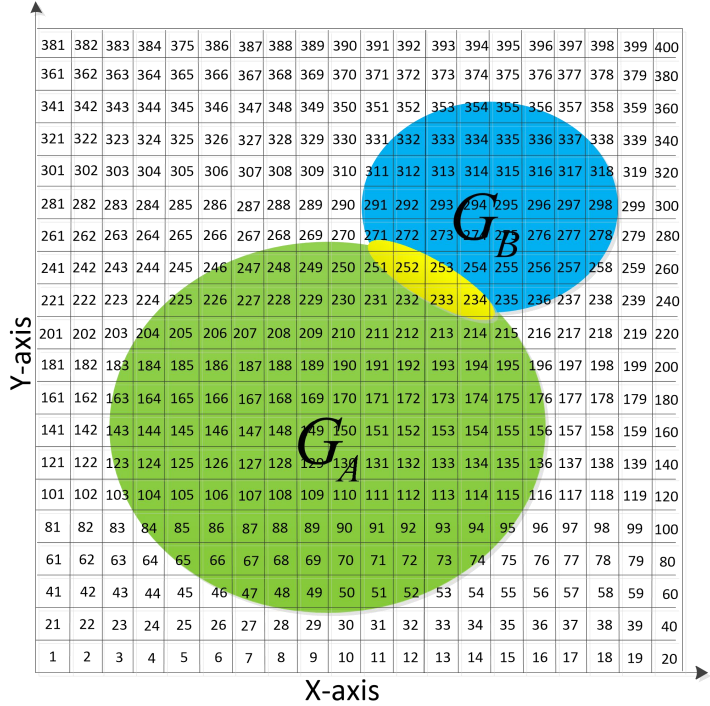

**Figure 1:** The illustration of partitioning and numbering the plane with $R$=400. The green (blue) part is Alice's graph $G_A$ (Bob' graph $G_B$), respectively, and the yellow part is the intersection area.

Through representing Alice's and Bob's private geometric graphs $G_A$, $G_B$ as the grid sets $S_A$, $S_B$), the PQGI protocol is defined as follows.

**Definition 1 (the PQGI protocol):** Alice and Bob encode their serial numbers of graph grids, i.e., $S_A = (a_0, a_1, \cdots, a_{M-1})$, $S_B = (b_0, b_1, \cdots, b_{N-1})$, into two initial states $|\psi_A\rangle = \frac{1}{\sqrt{M}} \sum_{i=0}^{M-1} |i\rangle \otimes |0\rangle^{\otimes r}$ and

 

$|\psi_B\rangle = \frac{1}{\sqrt{N}}\sum_{j=0}^{N-1}|j\rangle \otimes |0\rangle^{\otimes r}$ , respectively, here $r = \lceil \log R \rceil$ . After executing this protocol, they can obtain the result of whether the two graphs intersect without revealing their private information. To be specific, the PQGI protocol should guarantee the following privacy:

- **Alice's Privacy** Bob cannot learn any secret information about Alice's geometric graph without risking Alice's detection.

- **Bob's Privacy** Alice cannot get any secret information about Bob's geometric graph without risking Bob's detection.

## 3. The privacy-preserving quantum two-party geometric intersection protocol

Suppose Alice and Bob's private geometric graphs $G_A$ and $G_B$ are located on a unified plane, and the plane is uniformly divided into $R$ grids, here $R$ is a large enough integer that the whole plane can be represented by these grids with sufficient accuracy. Thus Alice's and Bob's graphs $G_A$, $G_B$ can be represented as the sets of grids: $S_A = (a_0, a_1, \cdots, a_{M-1})$, $S_B = (b_0, b_1, \cdots, b_{N-1})$, where $a_i, b_j$ are unique serial numbers in $[1, R]$, $0 \le i \le M-1$, $0 \le j \le N-1$. The detailed protocol is described in detail as follows (shown in Fig. 2).

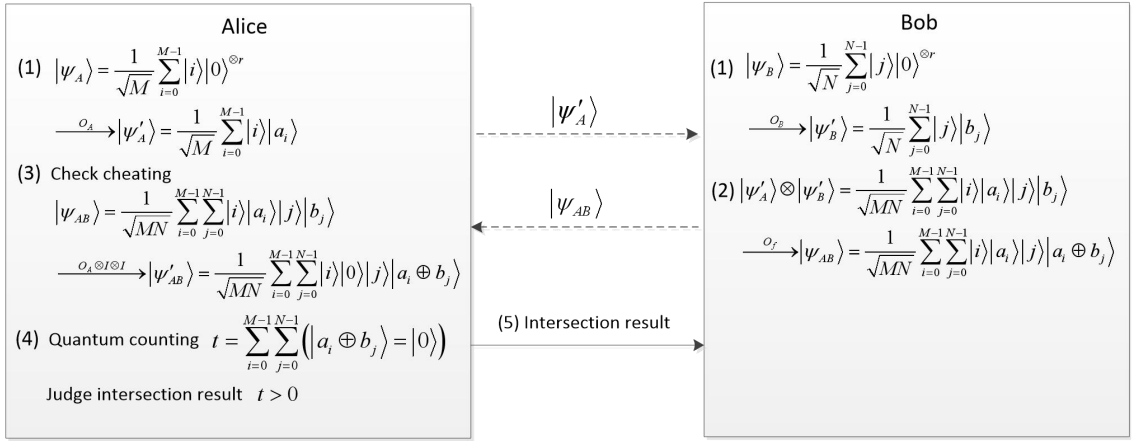

**Figure 2:** The procedure of the proposed PQGI protocol. The dotted (solid) line denotes the quantum (classic) channel.

1. Alice and Bob prepare the initial states $\frac{1}{\sqrt{M}}\sum_{i=0}^{M-1}|i\rangle_{A_a}|0\rangle_{D_a}^{\otimes r}$ , $\frac{1}{\sqrt{N}}\sum_{j=0}^{N-1}|j\rangle_{A_b}|0\rangle_{D_b}^{\otimes r}$ , respectively, where $0 \le i \le M-1$, $r = \lceil \log R \rceil$, $0 \le j \le N-1$. $A_a$ and $A_b$ denote Alice's ($m$-qubit) and Bob's ($n$-qubit) address qubits, while $D_a$ and $D_b$ represent Alice's and Bob's ($r$-qubit) data qubits, $m = \lceil \log M \rceil$, $n = \lceil \log N \rceil$. Then Alice and Bob apply the oracle operation $O_A$, $O_B$ on $|\psi_A\rangle$, $|\psi_B\rangle$ to encode their private elements of $S_A = (a_0, a_1, \cdots, a_{M-1})$, $S_B = (b_0, b_1, \cdots, b_{N-1})$ (shown in Fig. 3 and Fig.4) .

$$\frac{1}{\sqrt{M}}\sum_{i=0}^{M-1}|i\rangle \otimes |0\rangle^{\otimes r} \xrightarrow{O_A} \frac{1}{\sqrt{M}}\sum_{i=0}^{M-1}|i\rangle|a_i\rangle \qquad (1)$$

$$\frac{1}{\sqrt{N}}\sum_{j=0}^{N-1}|j\rangle\otimes|0\rangle^{\otimes r} \xrightarrow{\quad O_B \quad} \frac{1}{\sqrt{N}}\sum_{j=0}^{N-1}|j\rangle|b_j\rangle$$

(2)

Then Alice and Bob obtain the result states $|\psi'_A\rangle=\frac{1}{\sqrt{M}}\sum_{i=0}^{M-1}|i\rangle|a_i\rangle$ , $|\psi'_B\rangle=\frac{1}{\sqrt{N}}\sum_{j=0}^{N-1}|j\rangle|b_j\rangle$ ,

respectively, then Alice sends her state $|\psi'_A\rangle$ to Bob.

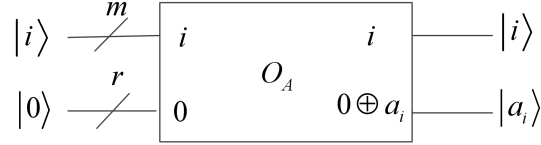

**Figure 3:** Schematic circuit of the oracle operation $O_A$ .

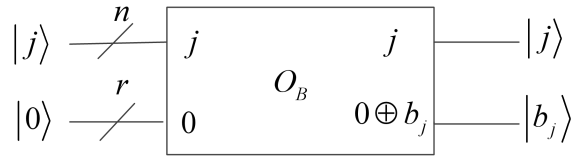

**Figure 4:** Schematic circuit of the oracle operation $O_B$ .

2. After receiving $|\psi'_A\rangle$ from Alice, Bob applies oracle operation $O_f$ on $|\psi'_A\rangle\otimes|\psi'_B\rangle$ (shown in Fig. 5), where $O_f$ works as follows:

$$\frac{1}{\sqrt{MN}}\sum_{i=0}^{M-1}\sum_{j=0}^{N-1}|i\rangle|a_i\rangle|j\rangle|b_j\rangle \xrightarrow{\quad O_f \quad} \frac{1}{\sqrt{MN}}\sum_{i=0}^{M-1}\sum_{j=0}^{N-1}|i\rangle|a_i\rangle|j\rangle|a_i\oplus b_j\rangle$$

(3)

Then he sends the result state $|\psi_{AB}\rangle=\frac{1}{\sqrt{MN}}\sum_{i=0}^{M-1}\sum_{j=0}^{N-1}|i\rangle|a_i\rangle|j\rangle|a_i\oplus b_j\rangle$ to Alice.

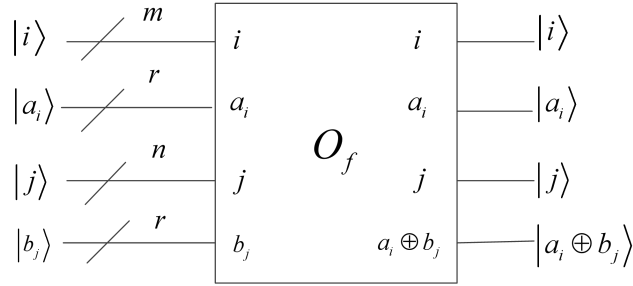

**Figure 5:** Schematic circuit of the oracle operation $O_f$ .

3. After receiving $|\psi_{AB}\rangle$ from Bob, Alice checks whether Bob cheated by applying oracle operation $O_A$ on the data qubits $D_a$ in $|\psi_{AB}\rangle$ as follow (shown in Fig. 6) :

$$\frac{1}{\sqrt{MN}}\sum_{i=0}^{M-1}\sum_{j=0}^{N-1}|i\rangle|a_i\rangle|j\rangle|a_i\oplus b_j\rangle \xrightarrow{\quad O_A\otimes I\otimes I \quad} \frac{1}{\sqrt{MN}}\sum_{i=0}^{M-1}\sum_{j=0}^{N-1}|i\rangle|0\rangle|j\rangle|a_i\oplus b_j\rangle$$

(4)

The result state is named $\left|\psi'_{AB}\right\rangle$, Alice then performs measurement on the data qubits $D_a$ in $\left|\psi'_{AB}\right\rangle$, if the measurement outcome turns to be $\left|0\right\rangle$, she can conclude that Bob has not cheated.

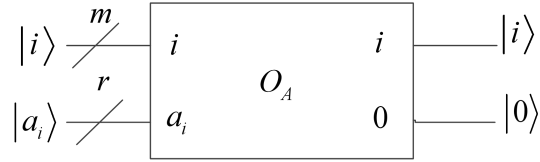

**Figure 6:** Schematic circuit of the oracle operation $O_A$.

4. After the cheating check, Alice executes the quantum counting algorithm [Brassard, HØyer and Tapp (1998)] on $\left|\psi'_{AB}\right\rangle$ to count $t = \sum_{i=0}^{M-1}\sum_{j=0}^{N-1}\left(\left|a_i\oplus b_j\right\rangle = \left|0\right\rangle\right)$, where $t$ is the number of states that $\left|a_i\oplus b_j\right\rangle$ equaling to $\left|0\right\rangle$ in $\left|\psi'_{AB}\right\rangle$ ( $i\in[0, M-1]$, $j\in[0, N-1]$ ). After executing the quantum algorithm, Bob obtains the result of $t$. Then Alice judges whether $G_A$ and $G_B$ intersect according to the value of $t$: if $t>0$, then it can be deduced that there exists $a_i = b_j$ for any $i$ and $j$, then get the conclusion that $G_A$ intersects with $G_B$, otherwise, $G_A$ and $G_B$ are not intersect.

5. Alice tells Bob the result of whether $G_A$ intersects with $G_B$.

## 3. Correctness, security and efficiency analysis

### 3.1. Correctness analysis

Without loss of generality, we suppose that Alice (Bob) has a private graph $G_A$ ( $G_B$ ), which is represented as $S_A = \{1, 2, 5, 6\}$ ($S_B = \{6, 7, 10, 11\}$), and thus $M$=4, $N$=4, $R$=16, $m = \lceil \log M \rceil = 2$, $n = \lceil \log N \rceil = 2$, $r = \lceil \log R \rceil = 4$. Alice and Bob want to determine whether there exists an intersection between $G_A$ and $G_B$ (shown in Fig. 7).

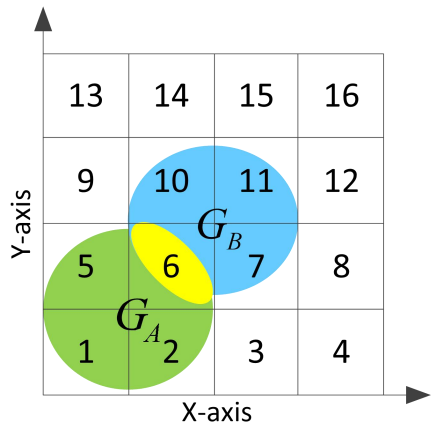

**Figure 7:** The example of the two intersecting geometric graphs $G_A$ and $G_B$.

In Step 1, Alice and Bob prepare two initial quantum states $|\psi_A\rangle$, $|\psi_B\rangle$ in the form of $\frac{1}{2}(|0\rangle+|1\rangle+|2\rangle+|3\rangle)\otimes|0\rangle$, and then apply oracle operation $O_A$, $O_B$ on them, the two result states $|\psi'_A\rangle$, $|\psi'_B\rangle$ are as follows:

$$
\begin{aligned}
|\psi'_A\rangle &= \frac{1}{2}\sum_{i=0}^{3}|i\rangle|a_i\rangle \\
&= \frac{1}{2}(|0\rangle|1\rangle+|1\rangle|2\rangle+|2\rangle|5\rangle+|3\rangle|6\rangle)
\end{aligned}
\tag{5}
$$

$$
\begin{aligned}
|\psi'_B\rangle &= \frac{1}{2}\sum_{j=0}^{3}|j\rangle|b_j\rangle \\
&= \frac{1}{2}(|0\rangle|6\rangle+|1\rangle|7\rangle+|2\rangle|10\rangle+|3\rangle|11\rangle)
\end{aligned}
\tag{6}
$$

Then Alice sends state $|\psi'_A\rangle$ to Bob. In step 2, Bob attaches $|\psi'_A\rangle$ with his state $|\psi'_B\rangle$, and generates the state $|\psi'_A\rangle\otimes|\psi'_B\rangle$ as below:

$$
\begin{aligned}
|\psi'_A\rangle\otimes|\psi'_B\rangle &= \frac{1}{4}\sum_{i=0}^{3}\sum_{j=0}^{3}|i\rangle|a_i\rangle|j\rangle|b_j\rangle \\
&= \frac{1}{4}(|0\rangle|1\rangle+|1\rangle|2\rangle+|2\rangle|5\rangle+|3\rangle|6\rangle)(|0\rangle|6\rangle+|1\rangle|7\rangle+|2\rangle|10\rangle+|3\rangle|11\rangle) \\
&= \frac{1}{4}(|0\rangle|1\rangle|0\rangle|6\rangle+|0\rangle|1\rangle|1\rangle|7\rangle+|0\rangle|1\rangle|2\rangle|10\rangle+|0\rangle|1\rangle|3\rangle|11\rangle \\
&\quad +|1\rangle|2\rangle|0\rangle|6\rangle+|1\rangle|2\rangle|1\rangle|7\rangle+|1\rangle|2\rangle|2\rangle|10\rangle+|1\rangle|2\rangle|3\rangle|11\rangle \\
&\quad +|2\rangle|5\rangle|0\rangle|6\rangle+|2\rangle|5\rangle|1\rangle|7\rangle+|2\rangle|5\rangle|2\rangle|10\rangle+|2\rangle|5\rangle|3\rangle|11\rangle \\
&\quad +|3\rangle|6\rangle|0\rangle|6\rangle+|3\rangle|6\rangle|1\rangle|7\rangle+|3\rangle|6\rangle|2\rangle|10\rangle+|3\rangle|6\rangle|3\rangle|11\rangle
\end{aligned}
\tag{7}
$$

Then Bob applies the oracle operation $O_f$ on state $|\psi'_A\rangle\otimes|\psi'_B\rangle$, and obtains the state $|\phi_{AB}\rangle$,

$$
\begin{aligned}
|\psi_{AB}\rangle &= \frac{1}{4}\sum_{i=0}^{3}\sum_{j=0}^{3}|i\rangle|a_i\rangle|j\rangle|a_i\oplus b_j\rangle \\
&= \frac{1}{4}(|0\rangle|1\rangle|0\rangle|7\rangle+|0\rangle|1\rangle|1\rangle|6\rangle+|0\rangle|1\rangle|2\rangle|11\rangle+|0\rangle|1\rangle|3\rangle|10\rangle \\
&\quad +|1\rangle|2\rangle|0\rangle|4\rangle+|1\rangle|2\rangle|1\rangle|5\rangle+|1\rangle|2\rangle|2\rangle|8\rangle+|1\rangle|2\rangle|3\rangle|9\rangle \\
&\quad +|2\rangle|5\rangle|0\rangle|3\rangle+|2\rangle|5\rangle|1\rangle|2\rangle+|2\rangle|5\rangle|2\rangle|15\rangle+|2\rangle|5\rangle|3\rangle|14\rangle \\
&\quad +|3\rangle|6\rangle|0\rangle|0\rangle+|3\rangle|6\rangle|1\rangle|1\rangle+|3\rangle|6\rangle|2\rangle|13\rangle+|3\rangle|6\rangle|3\rangle|13\rangle
\end{aligned}
\tag{8}
$$

Bob further sends the result state $|\psi_{AB}\rangle$ to Alice. Alice then applies the oracle operation $O_A$ on the data qubits $D_a$ of $|\psi_{AB}\rangle$ and obtains the result state $|\psi'_{AB}\rangle$ as follow :

$$
\begin{aligned}
|\psi'_{AB}\rangle &= \frac{1}{4}\sum_{i=0}^{3}\sum_{j=0}^{3}|i\rangle|0\rangle|j\rangle|a_i\oplus b_j\rangle \\
&= \frac{1}{4}(|0\rangle|0\rangle|0\rangle|7\rangle+|0\rangle|0\rangle|1\rangle|6\rangle+|0\rangle|0\rangle|2\rangle|11\rangle+|0\rangle|0\rangle|3\rangle|10\rangle \\
&\quad +|1\rangle|0\rangle|0\rangle|4\rangle+|1\rangle|0\rangle|1\rangle|5\rangle+|1\rangle|0\rangle|2\rangle|8\rangle+|1\rangle|0\rangle|3\rangle|9\rangle \\
&\quad +|2\rangle|0\rangle|0\rangle|3\rangle+|2\rangle|0\rangle|1\rangle|2\rangle+|2\rangle|0\rangle|2\rangle|15\rangle+|2\rangle|0\rangle|3\rangle|14\rangle \\
&\quad +|3\rangle|0\rangle|0\rangle|0\rangle+|3\rangle|0\rangle|1\rangle|1\rangle+|3\rangle|0\rangle|2\rangle|13\rangle+|3\rangle|0\rangle|3\rangle|13\rangle
\end{aligned}
\tag{9}
$$

Alice then performs measurement on the data qubits $D_a$ of $|\psi'_{AB}\rangle$, the measurement outcome turns to be $|0\rangle$, then she can conclude that Bob has not cheated. Then Alice executes the quantum counting algorithm on $|\psi'_{AB}\rangle$. Since the counting result $t = \sum_{i=0}^{3}\sum_{j=0}^{3}\left(\left|a_i \oplus b_j\right\rangle = |0\rangle\right) = 1 > 0$, thus graph $G_A$ intersects with graph $G_B$.

### 3.2. Security analysis

Now we discuss the security of our protocol. To realize such a secure PQGI protocol, two security requirements should be satisfied, that are Alice's privacy and Bob's privacy.

### 3.2.1 Alice's privacy

Suppose Bob wants to extract information about private graph $G_A$ (i.e., $a_i$ without affecting the final result of the protocol. If Bob performs the projective measurement on state $|\psi'_A\rangle = \frac{1}{\sqrt{M}}\sum_{i=0}^{M-1}|i\rangle|a_i\rangle$, he can randomly obtain one element $a_i$ from $|\psi'_A\rangle$. The state $|\psi'_A\rangle$ can also be represented by an ensemble $\varepsilon \equiv \{P_i, \rho(i)\}$, here $P_i = \frac{1}{M}$ is the probability that Bob obtains Alice's coordinates:

$$\rho = |\psi'_A\rangle = \frac{1}{\sqrt{M}}\sum_{i=0}^{M-1}|i\rangle|a_i\rangle. \tag{10}$$

$$\rho(i) = |a_i\rangle|i\rangle\langle i||\langle a_i|. \tag{11}$$

Here, we get the upper bound of information that Bob can get from Alice's coordinates is determined by the Holevo's bound [Holevo (2011)]:

$$H(B:A) \le S(\rho) - \frac{1}{M}\sum_{i=0}^{M-1}S(\rho(i)), \tag{12}$$

where $S(\rho)$ denotes the Von Neumann entropy of quantum state $\rho$, $H(A:B)$ means the information Bob can get about Alice's secret information, we have:

$$S(\rho) = S(\frac{1}{M}\sum_{i=1}^{M-1}|i\rangle|a_i\rangle\langle a_i||\langle i|) = \log(MR). \tag{13}$$

and $S(\rho(i)) = S\left(|i\rangle|a_i\rangle\langle a_i||\langle i|\right) = 0$, therefore:

$$H(B:A) \le \log(MR). \tag{14}$$

Then, Bob can only get coordinate information by measuring the state $\rho$. If Bob performs measurement on the state $\frac{1}{\sqrt{M}}\sum_{i=0}^{M-1}|i\rangle|a_i\rangle$, the state will collapse into one basis state, i.e., $|a_\tau\rangle$, $\tau \in [0, M-1]$ and the state $|\psi_{AB}\rangle$ will be changed to $|\tau\rangle \otimes |a_\tau\rangle \otimes \frac{1}{\sqrt{N}}\sum_{j=0}^{N-1}|j\rangle|a_i \oplus b_j\rangle$. In step 3 of our protocol, Alice checks whether Bob has cheated by applying oracle operation $O_A$ on the data qubits $D_a$ in $|\psi'_{AB}\rangle$ and then performs measurement on them. Since the measured data qubits $|a_\tau\rangle$ does not equal to $|0\rangle$, she can conclude Bob has cheated and aborts the protocol.

### 3.2.2 Bob's privacy

Suppose Alice wants to extract any information about private graph $G_B$ (i.e., $b_i$ without affecting the final result of the protocol. If Bob performs the projective measurement on state $\frac{1}{\sqrt{MN}} \sum_{i=0}^{M-1} \sum_{j=0}^{N-1} |i\rangle |a_i\rangle |j\rangle |a_i \oplus b_j\rangle$, he can randomly obtain one element, i.e., $a_i \oplus b_j$. However, the state Alice received is $|\psi_{AB}\rangle = \frac{1}{\sqrt{MN}} \sum_{i=0}^{M-1} \sum_{j=0}^{N-1} |i\rangle |a_i\rangle |j\rangle |a_i \oplus b_j\rangle$, and Alice does not know choose which base to measure and obtain $b_j$. On the other hand, the received information is in the form of $a_i \oplus b_j$, which means he even does not know which $a_i$ encodes the $b_j$, and therefore prevents his cheating on Bob's privacy.

### 3.3. Efficiency analysis

The communication cost is one of the key indicators of the efficiency for communication protocols. In order to analyze the efficiency of our PQGI protocol, we choose the classical PGI protocols [Atallah and Du (2001); Qin, Duan, Zhao et al. (2014)] as comparative references. In the Atallah et al.'s protocol, the participants send total $4M^2$ messages to Bob, here $M$ is the number of divided edges of the geometric graph, and each message requires $R$ bits. So the transmitted messages of their protocol are $4M^2 * R$, and their communication complexity is $O(M^2R)$. While in Qin et al.'s protocol, it requires to send $2(M^2 + N^2)$ messages, and each message requires $R$ bits. Here $M$ and $N$ are the number of curves from the edges of the geometry, and its communication complexity is $O((M^2 + N^2)R)$.

In our PQGI protocol, Alice sends a ($m+r$)-qubit states $|\psi'_A\rangle$ to Bob in Step 1, and then Bob sends a ($m+n+2r$)-qubit state $|\psi_{AB}\rangle$ to Alice in Step 3, where $m = \lceil \log M \rceil$, $n = \lceil \log N \rceil$ and $r = \lceil \log R \rceil$, thus the total transmitted messages of our protocol are $2\lceil \log M \rceil + \lceil \log N \rceil + 4\lceil \log R \rceil$ qubits. Thus our communication complexity is $O(\log MNR)$. Through the above calculations, we can get the results of the three protocols' communication complexity (see Tab. 1). Obviously, our protocol achieves a great reduction in the communication complexity aspect.

**Table 1:** Comparison among our protocol and the other PGI protocols

| Protocols | Transmitted messages | Communication complexity |
|---|---|---|
| Atallah et al.'s | $4M^2 * R$ | $O(M^2R)$ |
| Qin et al.'s | $2(M^2 + N^2)R$ | $O((M^2 + N^2)R)$ |
| Our | $2\lceil \log M \rceil + \lceil \log N \rceil + 4\lceil \log R \rceil$ | $O(\log MNR)$ |

## 4. Conclusion and discussion

In this paper, we present a novel quantum solution to two-party geometric intersection based on oracle and the quantum counting algorithm. The security of them is based on the quantum cryptography instead of difficulty assumptions of mathematical problem. Compared with the classical related protocols, our solution has the advantage of higher security and lower communication complexity. In addition, our proposed protocol can also be extended to some other complicated privacy-preserving computation problems, such as privacy-preserving database queries over cloud data [Cao, Wang, Li et al. (2014); Shen, Li, Li et al. (2017)], privacy-preserving set operations in cloud computing [Cao, Li, Dang et al. (2017); Zhuo, Jia, Guo et al. (2017)], and privacy-preserving reversible data hiding over encrypted image [Cao, Du, Wei et al. (2016)].

Furthermore, the method of the oracle operation applied in the presented protocols is general and can be employed to solve other similar privacy-preserving computation geometry protocols, which have the property that geometric graphs can be divided into small enough grids, such as privacy-preserving convex hull. However, how to extend our two party scenarios to the multi-party scenario, and the more complex situations such as geometric union is another problem. We would like to investigate the applications of quantum technologies in more kinds of privacy-preserving computational geometric protocols in the future.

**Acknowledgement**: The authors would like to thank the anonymous reviewers and editor for their comments that improved the quality of this paper. This work is supported by the National Nature Science Foundation of China (Grant Nos. 61502101 and 61501247), the Natural Science Foundation of Jiangsu Province, China (Grant No. BK20171458), the Six Talent Peaks Project of Jiangsu Province, China (Grant No. 2015-XXRJ-013), the Natural science Foundation for colleges and universities of Jiangsu Province, China (Grant No. 16KJB520030), the Research Innovation Program for College Graduates of Jiangsu Province, China (Grant No. KYCX17_0902), the Practice Innovation Training Program Projects for the Jiangsu College Students (Grant No. 201810300016Z), and the Priority Academic Program Development of Jiangsu Higher Education Institutions (PAPD).

**References**

**Assaf, B.; Gawrilow, E.; Herr, K.; Joswig, M.; Lorenz, B.; Paffenholz, A.;** et al. (2017): Computing convex hulls and counting integer points with polymake. *Mathematical Programming Computation*, vol. 9, no. 1, pp. 1-38.

**Atallah, M. J.; Du, W.** (2001): Secure multi-party computational geometry. In *Proceedings of the 7th International Workshop on Algorithms and Data Structures*, pp. 165-179.

**Bennett, C. H.; Brassard, G.** (1984): Quantum cryptography: public key distribution and coin tossing. In *International conference on Computers, Systems & Signal Processing*, pp.10-12.

**Brassard, G.; HØyer, P.; Tapp, A.** (1998): Quantum counting. In *International Colloquium on Automata, Languages, and Programming*, pp. 820-831.

**Erlebach, T.; Jansen, K.; Seidel, E.** (2005): Polynomial-time approximation schemes for geometric intersection graphs. *SIAM Journal on Computing*, vol. 34, no. 6, pp. 1302-1323.

**Cao, N.; Wang, C.; Li, M.; Ren, K; Lou, W. J.** (2014) Privacy-preserving multi-keyword ranked search over encrypted cloud data, *IEEE Transactions on Parallel and Distributed Systems*, vol. 25, no. 1, pp. 497-503.

**Cao, X. F.; Li, H.; Dang, L. J.; Yin, L.** (2017) A two-party privacy preserving set intersection protocol against malicious users in cloud computing, *Computer Standards & Interfaces*, vol. 54, no. 1, pp. 41-45.

**Cao, X. H.; Du, L; Wei, X. X.; Meng, D.; Guo, X. J.** (2016) High capacity reversible data hiding in encrypted images by patch-level sparse representation, *IEEE Transactions on Cybernetics*, vol. 46, no. 5, pp. 1132-1143.

**Grover, L. K.** (1996): A fast quantum mechanical algorithm for database search. In *Proceedings of the Twenty-eighth Annual ACM Symposium on Theory of Computing*, pp. 212-219.

**Holevo, A. S.** (2011): *Probabilistic and Statistical Aspects of Quantum Theory*, Springer Science & Business Media, USA.

**Huang, L.; Luo, Y.; Wang, Q.** (2008): Privacy-preserving protocols for finding the convex hulls. In *2008 Third International Conference on Availability, Reliability and Security (ARES)*, pp. 727-732.

**Li, C.; Ni, R.** (2002): Derivatives of generalized distance functions and existence of generalized nearest points. *Journal of Approximation Theory*, vol. 115, no. 1, pp. 44-55.

**Li, S.; Wu, C.; Wang, D.; Dai, Y.** (2014): Secure multiparty computation of solid geometric problems and their applications. *Information Sciences*, vol. 282, pp. 401-413.

**Liu, W. J.; Chen, H. W.; Ma, T. H.; Li, Z. Q.; Liu, Z. H.; Hu, W. B.** (2009): An efficient deterministic secure quantum communication scheme based on cluster states and identity authentication. *Chinese Physics B*, vol. 18, no. 10, pp. 4105-4109.

**Liu, W. J.; Liu, C.; Liu, Z. H.; Liu, J. F.; Geng, H. T.** (2014): Same initial states attack in Yang *et* quantum private comparison protocol and the improvement. *International Journal of Theoretical Physics*, vol. 53, no. 1, pp. 271-276.

**Liu, W. J.; Liu, C.; Chen, H. W.; Li, Z. Q.; Liu, Z. H.** (2014): Cryptanalysis and improvement of quantum private comparison protocol based on bell entangled states. *Communications in Theoretical Physics*, vol. 62, no. 8, pp. 210-214.

**Liu, W. J.; Liu, C.; Wang, H. B.; Liu, J. F.; Wang, F.; Yuan, X. M.** (2014): Secure quantum private comparison of equality based on asymmetric w state. *International Journal of Theoretical Physics*, vol. 53, no. 6, pp. 1804-1813.

**Liu, W. J.; Wang, F.; Ji, S.; Qu, Z. G.; Wang, X. J.** (2014): Attacks and improvement of quantum sealed-bid auction with epr pairs. *Communications in Theoretical Physics*, vol. 61, no. 6, p. 686.

**Liu, W. J., Chen, Z. F., Liu, C., Zheng, Y.** (2015)**:** Improved deterministic N-to-one joint remote preparation of an arbitrary qubit via EPR pairs. *International Journal of Theoretical Physics*, vol. 54, no. 2, pp. 472-483.

**Liu, W. J.; Wang, H. B.; Yuan, G. L.; Xu, Y.; Chen, Z. Y.; An, X. X.** et al. (2016): Multiparty quantum sealed-bid auction using single photons as message carrier. *Quantum Information Processing*, vol. 15, no. 2, pp. 869-879.

**Liu, W. J.; Chen, Z. Y.; Ji, S.; Wang, H. B.; Zhang, J.** (2017): Multi-party semi-quantum key agreement with delegating quantum computation. *International Journal of Theoretical Physics*, vol. 56, no. 10, pp. 3164-3174.

**Liu, W. J.; Xu, Y.; Yang, C. N.; Gao, P. P.; Yu, W. B.** (2018): An efficient and secure arbitrary n-party quantum key agreement protocol using bell states. *International Journal of Theoretical Physics*, vol. 57, no. 1, pp. 195-207.

**Liu, Z. H.; Chen, H. W.; Liu, W. J.** (2016): Information leakage problem in high-capacity quantum secure communication with authentication using Einstein-Podolsky-Rosen pairs. *Chinese Physics Letters*, vol. 33, no. 7, p. 070305.

**Liu, Z. H.; Chen, H. W.** (2016): Cryptanalysis and improvement of quantum broadcast communication and authentication protocol with a quantum one-time pad. *Chinese Physics B*, vol. 25, no. 8, p. 080308.

**Löffler, M.; van Kreveld, M.** (2010): Largest and smallest convex hulls for imprecise points. *Algorithmica*, vol. 56, no. 2, p. 235

**Luo, Y. L.; Huang, L. S.; Zhong, H.** (2007): Secure two-party point-circle inclusion problem. *Journal of Computer Science and Technology*, vol. 22, no. 1, pp. 88-91.

**Naseri, M.** (2009): Secure quantum sealed-bid auction. *Optics Communications*, vol. 282, no. 9, pp. 1939-1943.

**Oleshchuk, V. A.; Zadorozhny, V.** (2007): Secure multi-party computations and privacy preservation: Results and open problems. *Telektronikk*, vol. 103, no. 2, p. 20.

**Pawlik, A.; Kozik, J.; Krawczyk, T.; Laso´ n, M.; Micek, P.; Trotter, W. T.;** et al.(2013): Triangle-free geometric intersection graphs with large chromatic number. *Discrete & Computational Geometry*, vol. 50, no. 3, pp. 714-726.

**Preparata, F. P.; Shamos, M. I.** (2012): *Computational Geometry: An Introduction*. Springer Science & Business Media, USA.

**Qin, J.; Duan, H.; Zhao, H.; Hu, J.** (2014): A new lagrange solution to the privacy-preserving general geometric intersection problem. *Journal of Network and Computer Applications*, vol. 46, pp. 94-99.

**Qu, Z. G.; Wu, S. Y.; Wang, M. M.; Sun, L.; Wang, X. J.** (2017): Effect of quantum noise on deterministic remote state preparation of an arbitrary two-particle state via various quantum entangled channels, *Quantum Information Processing*, vol. 16, no. 306, pp. 1-25.

**Shen, Y.; Li, W. Y.; Li, L.; Huang, L. S.** (2017) Achieving fully privacy-preserving private range queries over outsourced cloud data, *Pervasive and Mobile Computing*, vol. 39, pp. 36-51.

**Shor, P. W.** (1994): Algorithms for quantum computation: Discrete logarithms and factoring. In *Proceedings of the 35th Annual Symposium on Foundations of Computer Science*, pp. 124-134.

**Tao, Y.; Yi, K.; Sheng, C.; Kalnis, P.** (2010): Efficient and accurate nearest neighbor and closest pair search in high-dimensional space. *ACM Transactions on Database Systems*, vol. 35, no. 3, p. 20.

**Troncoso-Pastoriza, J. R.; Katzenbeisser, S.; Celik, M.; Lemma, A.** (2007): A secure multidimensional point inclusion protocol. In *Proceedings of the 9th Workshop on Multi-media & Security*, pp. 109-120.

**Zhuo, G. Q; Jia, Q.; Guo, L. K.; Li, M.; Li, P.** (2017) Privacy-preserving verifiable set operation in big data for cloud-assisted mobile crowdsourcing, *IEEE Internet of Things Journal*, vol. 4, no. 2, pp. 572 - 582.